\documentclass[11pt,letterpaper,superscriptaddress]{JHEP3}
\usepackage{graphicx}
\usepackage{amssymb, amsmath,amsfonts,amstext,graphics}
\usepackage{epstopdf}
\usepackage{cite}
\input epsf.tex
\newcommand{\lang}{\mathcal{L}}

\newcommand{\pd}[2]{\frac{\partial #1}{\partial #2}}

\newcommand{\beq}{\begin{align}}
\newcommand{\eeq}{\end{align}}

\def\ltap{\ \raise.3ex\hbox{$<$\kern-.75em\lower1ex\hbox{$\sim$}}\ }
\def\gtap{\ \raise.3ex\hbox{$>$\kern-.75em\lower1ex\hbox{$\sim$}}\ }
\newcommand{\gsim}{\lower.7ex\hbox{$\;\stackrel{\textstyle>}{\sim}\;$}}
\newcommand{\lsim}{\lower.7ex\hbox{$\;\stackrel{\textstyle<}{\sim}\;$}}

\newcommand{\CP}{\mathcal{P}}

\def\CO{{\cal O}}

\def\tr{{\rm\ Tr}}
\def\CO{{\cal O}}

\def\be{\begin{equation}}
\def\ee{\end{equation}}
\def\bea{\begin{align}}
\def\eea{\end{align}}

\newcommand{\vev}[1]{ \left\langle {#1} \right\rangle }

\newcommand{\tev}{{\rm TeV}}

\newcommand{\TeV}{\,\mathrm{TeV}}
\newcommand{\GeV}{\,\mathrm{GeV}}

\def\unit{\relax{\rm 1\kern-.26em I}}
\newcommand{\half}{{\frac{1}{2}}}

\newcommand{\SUSY}{{\text{SUSY}}}

\newcommand{\Tr}{{\text{ Tr }}}

\title{ISS-flation}
\author{Nathaniel J. Craig$^{a,b}$\\
$^a$ 
Theory Group,\\ 
Stanford Linear Accelerator Center,\\
Menlo Park, CA 94025\\
$^b$ Institute for Theoretical Physics,\\ 
Stanford University,\\ 
Stanford, CA 94306}
\abstract{
Inflation may occur while rolling into the metastable supersymmetry-breaking vacuum of massive supersymmetric QCD. We explore the range of parameters in which slow-roll inflation and long-lived metastable supersymmetry breaking may be simultaneously realized. The end of slow-roll inflation in this context coincides with the spontaneous breaking of a global symmetry, which may give rise to significant curvature perturbations via inhomogenous preheating. Such spontaneous symmetry breaking at the end of inflation may give rise to observable non-gaussianities, distinguishing this scenario from more conventional models of supersymmetric hybrid inflation.}

\begin{document}

\section{Introduction}

Spontaneously broken supersymmetry (SUSY) has long been one of the most attractive possibilities for physics beyond the Standard Model \cite{Dimopoulos:1981zb}. Considerable effort has been devoted to elucidating dynamical mechanisms of SUSY-breaking, in which strong-coupling effects give rise to a naturally small SUSY-breaking scale \cite{Witten:1981nf}. The SUSY-breaking vacua of such theories need not be global minima of the potential, and metastability of phenomenologically-viable vacua appears to be generic when embedding the MSSM into a larger setting. Indeed, the study of metastable dynamical SUSY-breaking has recently undergone something of a renaissance, catalyzed by the observation that massive supersymmetric QCD (SQCD) possesses SUSY-breaking local minima whose lifetimes can be longer than the present age of the universe \cite{Intriligator:2006dd} (for an excellent review, see \cite{Intriligator:2007cp}). The simplicity and genericness of such theories makes them particularly well-suited to supersymmetric model-building.

Far from serving only as a possible resolution of the hierarchy problem, spontaneously-broken supersymmetry emerges frequently in inflationary settings as well \cite{Guth:1980zm, Linde:1981mu, Albrecht:1982wi, Linde:1983gd}. There exist a plethora of possible models aimed at realizing inflation in a natural context \cite{Lyth:1998xn}, many of which exploit  supersymmetry and supersymmetry breaking \cite{Kinney:1997hm, Kinney:1998dv, Dvali:1994ms, Dimopoulos:1997fv, Copeland:1994vg}. Based on the moderate success of such theories, it is tempting to suppose that inflation may be built into a supersymmetry-breaking sector. Indeed, it has already been demonstrated that inflation may arise in simple strongly-coupled gauge theories \cite{Dimopoulos:1997fv}. This suggests that it may be possible to inflate while rolling into the metastable vacuum of massive SQCD from sub-Planckian initial field values. The appeal of successful inflation in this scenario is twofold: first, it provides a generic and technically natural means of simultaneously realizing inflation and supersymmetry breaking; second, it furnishes a robust UV completion for the inflationary sector. Moreover, the spontaneous global symmetry-breaking that accompanies the end of inflation in these models may give rise to additional curvature perturbations exhibiting observable non-gaussianities. In this note we address the prospects for, and signatures of, slow-roll inflation in models with metastable dynamical supersymmetry breaking.

The organization of the paper is as follows.  In Sec.~\ref{sec:hybrid} we review the traditional scenario of supersymmetric hybrid inflation proposed in \cite{Dvali:1994ms}, wherein one-loop corrections drive inflation along a flat direction of a theory with a relatively simple superpotential. In Sec.~\ref{sec:hybridSQCD} we review the means by which supersymmetric hybrid inflation may be realized in strongly-coupled gauge theories \cite{Dimopoulos:1997fv}. Subsequently, in Sec.~\ref{sec:ISS} we turn to the Intriligator, Seiberg, and Shih (ISS) model \cite{Intriligator:2006dd}  of metastable supersymmetry breaking in supersymmetric QCD and explore its implications. In the vicinity of the supersymmetry breaking metastable vacuum, features of the ISS model are reminiscent of supersymmetric hybrid inflation. With this motivation, in Sec.~\ref{sec:metastable} we explore the prospects for, and constraints on, inflation while rolling into the metastable supersymmetry breaking vacuum of the ISS model; in Sec.~\ref{sec:predictions} we discuss the subsequent inflationary predictions assuming the inflaton is primarily responsible for the primordial curvature perturbation. However, the spontaneous symmetry breaking that accompanies the end of slow-roll inflation may give rise to additional curvature perturbations through inhomogenous preheating \cite{Kolb:2004jm}. In Sec.~\ref{sec:broken} we explore this possibility and its consequences, including the prospects for observable non-gaussianities. In Sec.~\ref{sec:conclusions} we conclude the analysis and consider directions for further work.


\section{Supersymmetric Hybrid Inflation}\label{sec:hybrid}

Hybrid inflation provides a compelling alternative to traditional chaotic inflation, in that a second non-inflationary field provides the bulk of the inflationary energy density and a natural end to inflation without recourse to trans-Planckian field values \cite{Linde:1991km, Linde:1993cn}. Such theories were found to arise naturally in the context of supersymmetry \cite{Copeland:1994vg}. One particularly effective model, due to \cite{Dvali:1994ms}, obtains the scalar potential of hybrid inflation from a simple globally-supersymmetric superpotential involving three chiral superfields: a superfield $\psi$ transforming as a fundamental of the gauge symmetry; a superfield $\psi^c$ transforming as an anti-fundamental; and a gauge singlet superfield $X.$ The superpotential of this theory is
\begin{align}
W_X = h \psi^c X \psi - \mu^2 X,
\label{eqn:hybridW}
\end{align}
which exhibits an unbroken R symmetry. 
In the D-flat directions, the scalar potential of the theory obtains the form \footnote{We will henceforth abuse notation by using the same notation for superfields and their scalar components.}
\begin{align}
V = h^2 |X|^2 (|\psi^c|^2 + |\psi|^2) + |h \psi^c \psi - \mu^2|^2 .
\end{align}
For $X \leq X_c = \mu/\sqrt{h},$ the minimum for the $\psi, \psi^c$ is at $\vev{\psi, \psi^c} = \mu/ \sqrt{h},$ and there exists a supersymmetric vacuum at 
\begin{align}
\vev{X} = 0 \hspace{1cm}
\vev{\psi, \psi^c} = \mu/ \sqrt{h}.
\end{align}
However, for $X > X_c,$ the minimum is at $\vev{\psi, \psi^c} = 0,$ thanks to the effective mass term arising from $\vev{X}.$ 

\subsection{Rolling from supersymmetry-breaking}

For $X > X_c,$ there arises a $\mu^4$ contribution to the vacuum energy density that drives inflation; inflation ends when $X < X_c$ and the scalars $\psi, \psi^c$ roll off into the supersymmetric vacuum. At tree level, there's nothing to drive $X$ to its minimum and end inflation; an adequate potential is obtained only when corrections at one loop are taken into consideration. Indeed, when $X > X_c$ the $F$-term for $X$ is nonzero and supersymmetry is broken; the scalar potential accumulates one-loop corrections of the form 
\begin{align}
V_1(X) = \sum_{i} \frac{(-1)^F}{64 \pi^2} M_i(X)^4 \log \left( \frac{M_i(X)^2}{\Lambda^2} \right)
\end{align}
where $\Lambda$ is a cutoff scale. The contributions to the effective potential from $X > X_c$ come from splittings in the $\psi, \psi^c$ superfields; the $\psi$ scalars have masses $m_s^2 = h^2 X^2 \pm h \mu^2,$ while the $\psi$ fermions have masses $ m_f^2 = h^2 X^2.$ The one-loop effective potential along the inflationary trajectory is then given by
\begin{align}
V_{eff}(X) = \mu^4 + \frac{N_c h^2}{32 \pi^2} \left[ 2 \mu^4 \log \left( \frac{h^2 |X|^2}{\Lambda^2} \right) + (h X^2 - \mu^2)^2 \log \left( 1 - \frac{\mu^2}{h X^2} \right) \right. \\
\left.  + ( h X^2 + \mu^2)^2 \log \left( 1 + \frac{\mu^2}{h X^2} \right) \right] \nonumber.
\end{align}
This contribution from the one-loop effective potential drives $X$ to the origin. When $X \simeq X_c,$ the $\psi, \psi^c$ become tachyonic and roll off to their supersymmetric minima. Below $X_c,$ the inflaton $X$ is efficiently driven to the origin by the  $\psi, \psi^c$ vevs. Inflation ends as the $\psi, \psi^c$ vevs cancel the effective contribution to the vacuum energy density and the slow-roll conditions are violated. Defining, for convenience, the parametrization $X = x X_c,$ i.e., $X = \frac{\mu}{\sqrt{h}} x,$ the effective potential is of the form
\begin{align}
V_{eff}(x) = \mu^4 + \frac{N_c h^2}{32 \pi^2} \left[ 2 \mu^4 \log \left( \frac{h x^2 \mu^2}{\Lambda^2} \right) + (x^2 \mu^2 - \mu^2)^2 \log \left( 1 - \frac{1}{x^2} \right) \right. \label{eqn:veff}\\
\left. + (x^2 \mu^2 + \mu^2)^2 \log \left( 1 + \frac{1}{x^2} \right) \right]. \nonumber
\end{align}
The slow-roll parameters in this scenario are (in reduced Planck units $M_P = 1/\sqrt{8 \pi G} = 2.4 \times 10^{18}$ GeV)
\begin{align}
\epsilon &= \frac{M_P^2}{2} \left( \frac{V'}{V} \right)^2 \simeq \frac{h^5 N_c^2 M_P^2}{128 \pi^4 \mu^2} x^2 \left[ (x^2 - 1) \log \left( 1 - \frac{1}{x^2} \right) + (x^2 + 1) \log \left( 1 + \frac{1}{x^2} \right) \right]^2 \\
\eta &= M_P^2 \frac{V''}{V} \simeq \frac{h^3 N_c M_P^2}{8 \pi^2 \mu^2} \left[ (3x^2 - 1) \log \left( 1 - \frac{1}{x^2} \right) + (3 x^2 + 1 ) \log \left( 1 + \frac{1}{x^2} \right) \right].
\end{align}
The slow-roll conditions $\eta \ll 1, \epsilon \ll 1$ are generally satisfied until $x \sim 1,$ where both $\epsilon, \eta$ grow rapidly and inflation comes to an end. Notice that $|\epsilon| \ll |\eta|$ and generally $\eta < 0,$ thereby guaranteeing a suitably red spectrum.

\subsection{Inflationary predictions for supersymmetric hybrid inflation}

The initial displacement $X_e$ required to obtain a sufficient number of e-foldings, $N_e \sim 54 \pm 7,$ is given by
\begin{align}
N_e \simeq \frac{1}{M_P^2} \int_{X_c}^{X_e} \left( \frac{V}{V'} \right) d X \simeq \frac{4 \pi^2}{N_c} \left( \frac{\mu}{M_P} \right)^2 \frac{(x_e^2-1)}{h^3}.
\end{align}
This suggests the field value of $X$ at $N_e$ e-foldings prior to the end of inflation is 
\begin{align}
X_e \simeq \frac{\sqrt{N_e N_c}}{2 \pi} h M_P.
\end{align}
This corresponds to a naturally sub-Planckian initial condition, provided $h \sqrt{N_c} \lsim 1.$

It is natural to ask whether such a theory is compatible with observation. The vacuum fluctuation of the inflaton generates a time-independent curvature perturbation $\zeta$ with spectrum \cite{Guth:1982ec, Hawking:1982cz, Starobinsky:1982ee, Bardeen:1983qw}
\begin{align}
\CP_{\zeta}^{1/2} \simeq \frac{V}{V'} \left( \frac{H}{2 \pi M_P^2} \right) \simeq \frac{1}{\sqrt{2 \epsilon}} \left( \frac{H}{2 \pi M_P} \right).
\end{align}
With $H = \sqrt{V_0/3 M_P^2} = \frac{h \mu^2}{\sqrt{3} M_P},$ this corresponds to 
\begin{align}
\CP_{\zeta}^{1/2} \simeq  \sqrt{\frac{4 N_e}{3 N_c}} \left( \frac{\mu}{M_P} \right)^{2}.
\end{align}
The WMAP normalization is $\CP_{\zeta}^{1/2} = 4.86 \times 10^{-5},$ as taken at the comoving scale $k_0 = 0.002 $ Mpc$^{-1}.$ Matching the observed curvature perturbation thus entails
\begin{align}
\frac{\mu}{M_P} \sim  2  \times 10^{-3} \cdot N_c^{1/4}
\end{align}
and hence the assumption of GUT-scale $\mu \sim 10^{15}$ GeV.

The spectral index, for $\epsilon \ll \eta,$ is given simply by 
\begin{align}
n_s \approx 1 - 2 |\eta| \approx 1 - \frac{1}{N_e} \approx 0.98,
\end{align}
somewhat higher than the central value $n_s \approx 0.95$ given by WMAP data \cite{Spergel:2006hy}. The tensor-to-scalar ratio is negligible, $r \lsim 10^{-4}, $ as is the running in the spectral index, $d n_s/d \log k \lsim 10^{-3}.$

Taken together, this simple scenario of globally supersymmetric hybrid inflation accumulates a number of successes: reasonable correspondence with the observed spectral index, sub-Planckian initial conditions, ready provision of $N_e \sim 54$ e-foldings, and a graceful exit from inflation. Moreover, as we shall see in subsequent sections, the requisite superpotential arises naturally in the context of supersymmetric gauge theories.

\section{Supersymmetric Hybrid Inflation from Strongly Coupled Theories}\label{sec:hybridSQCD}

Let us now consider scenarios wherein the superpotential (\ref{eqn:hybridW}) is generated in the context of strongly-coupled supersymmetric gauge theories, as was demonstrated in \cite{Dimopoulos:1997fv}. Consider an $SU(2)$ gauge theory whose matter content consists of four doublet chiral superfields $Q_I, \bar Q^J$ (with flavor indices $I,J = 1,2$) and a singlet superfield $S,$ with the classical superpotential
\begin{align}
W_{cl} = g S (Q_1 \bar Q_1 + Q_2 \bar Q_2)
\end{align}
with coupling constant $g.$ In the classical theory without any superpotential, the moduli space of D-flat directions is parametrized by the following $SU(2)$ invariants:
\begin{align}
S, \hspace{1cm} M_I^J = Q_I \bar Q^J, \hspace{1cm} B = \epsilon^{IJ} Q_I Q_J, \hspace{1cm} \bar{B} = \epsilon_{IJ} \bar Q^I \bar Q^J
\end{align}
These invariants are subject to the constraint
\begin{align}
\det M - \bar B B = 0.
\end{align}
With the superpotential turned on, the classical moduli space consists of two branches -- one with $S \neq 0, M = B = \bar B = 0,$ where the quarks are massive and the gauge symmetry is unbroken; and one with $S = 0,$ where the mesons and baryons satisfy the above constraint and the gauge group is broken. 

It is particularly interesting to consider the $S \neq 0$ branch, where a non-zero vacuum energy may drive inflation. For $S \neq 0,$ far from the origin, the `quarks' $Q, \bar Q$ become massive and decouple; the theory consists of a free singlet $S$ and pure supersymmetric $SU(2).$ The $SU(2)$ sector has an effective scale $\Lambda_L$ given by the one-loop matching to the quarks at the scale $\Lambda$ of the original theory,
\begin{align}
\Lambda_L^3 = g S \Lambda^2.
\end{align}
In the pure $SU(2)$ sector, gaugino condensation generates an effective superpotential
\begin{align}
W_{eff} = \Lambda_L^3 = g S \Lambda^2
\end{align}
This effective superpotential gives rise to a non-zero $F$-term, $F_S = g \Lambda^2,$ and supersymmetry is broken with a vacuum energy density $\sim g^2 \Lambda^4.$

While this is a convenient heuristic, it is useful to consider the full theory in detail. The quantum modified constraint for the confined theory is given by
\begin{align}
\det M - \bar B B - \Lambda^4 = 0,
\end{align}
so that the full quantum superpotential is given by
\begin{align}
W = A (\det M - \bar B B - \Lambda^4 ) + g S \tr M.
\end{align}
For $S \neq 0,$ the $F$-terms are 
\begin{align}
F_A &= \det M - B \bar B - \Lambda^4 \nonumber \\
F_{M_{I}^{J}} &= A \epsilon_{IK} \epsilon^{JL} M_{L}^{K} + g \delta_I^J S \nonumber \\
F_{S} &= g \tr M \\
F_{B} &= A \bar B \nonumber \\
F_{\bar B} &= A B \nonumber
\end{align}
 For $S \neq 0,$ all $F$-terms save $F_S$ may be set to zero via
 \begin{align}
 \bar B = B = 0, \hspace{5mm} M_I^J = \delta_I^J \Lambda^2, \hspace{5mm} A \Lambda^2/g = -S, \hspace{5mm} F_S = g \Lambda^2,
 \end{align}
which defines a natural D-flat trajectory for the rolling of $S.$

The inflationary behavior of the theory maps onto that of conventional supersymmetric hybrid inflation. For $S > S_c,$ the quarks are integrated out, providing a non-zero vacuum energy $\sim g^2 \Lambda^4$. The one-loop effective potential for $S$ drives the inflaton along its D-flat trajectory towards $S_c.$ Inflationary predictions are the same as before, albeit without the need to introduce any additional dimensionful parameters. By exploiting the dynamics of strongly-coupled theories, the hybrid inflation mass scale is set by the strong coupling scale $\Lambda.$

\section{Metastable SUSY breaking in Supersymmetric QCD}\label{sec:ISS}

In the preceding sections, we have seen how slow-roll inflation may arise naturally in the context of supersymmetric gauge theories. Consider now the case of massive supersymmetric QCD (SQCD). In \cite{Intriligator:2006dd} it was shown that for sufficiently small mass of the electric quarks in  $SU(N_c)$ SQCD, there is a long lived metastable vacuum in which supersymmetry is broken. Given the close resemblance of the ISS model  to the inflationary theories considered above, it is tempting to consider scenarios in which inflation occurs during the transition to a metastable SUSY-breaking vacuum.  In this section we shall briefly review the key features of the ISS model before turning to inflationary dynamics in subsequent sections.

\subsection{Microscopic (electric) theory}
 
In the ultraviolet, the microscopic (electric) ISS theory consists of asymptotically free $\mathcal{N} =1$ supersymmetric $SU(N_c)$ QCD with $N_f$ massive flavors $Q, Q^c$; the theory exhibits an $SU(N_f)_L\times SU(N_f)_R$ approximate global flavor symmetry. The quarks transform under the symmetries of the theory as
\begin{align}
Q \sim ( \square_{N_c}, \square_{N_{F\hspace{0.02in}L}})
\hspace{0.2in}
Q^c \sim ( \overline{\square}_{N_c}, \overline{\square}_{N_{F\hspace{0.02in}R}}).
\end{align}
The nonzero quark masses break the global symmetry to $SU(N_f)_{L} \times SU(N_f)_R \rightarrow SU(N_f)$ with a superpotential
\begin{align}
W_e=m \tr QQ^c.
\label{eqn:esuperpot}
\end{align}
For the sake of simplicity, the masses $m$ are taken to be degenerate. This theory with small $m$ is technically natural.

\subsection{Macroscopic (magnetic) theory}

This electric theory becomes strongly coupled at a scale $\Lambda$. In order to retain full control of the dynamics, we assume $m \ll \Lambda.$  Below this strong-coupling scale  the system may be described by an IR-free dual gauge theory provided $N_c < N_f  
< \frac{3}{2}N_c$ \cite{Seiberg:1994pq}.  The dual macroscopic (magnetic) theory is an $SU(N)$ gauge theory (where $N = N_f - N_c$) with $N_f$ magnetic `quarks' $q$  and $q^c.$ This dual theory possesses a Landau pole at a scale $\Lambda$ and runs free in the IR.\footnote{In principle the strong-coupling scale $\tilde \Lambda$ of the IR theory is different from the strong-coupling scale $\Lambda$ of the UV theory, related only through an intermediate scale $\hat \Lambda$ by the relation $\Lambda^{3 N_c - N_f} \tilde{\Lambda}^{3(N_f - N_c) - N_f} = (-1)^{N_f - N_c} \hat{\Lambda}^{N_f},$ with $\tilde \Lambda, \hat \Lambda$ not uniquely determined by the content of the electric theory. Naturalness and conceptual simplicity, however, suggest $\Lambda \sim \tilde{\Lambda},$ which will remain our convention throughout.}   The magnetic quarks  have the same approximate $SU(N_f)_L \times SU(N_f)_R$ flavor symmetry as the electric theory, transforming under the symmetries of the magnetic theory as
\begin{align}
q \sim ( \square_{N}, \overline{\square}_{N_{F\hspace{0.01in}L}})
\hspace{0.2in}
q^c \sim ( \overline{\square}_{N}, \square_{N_{F\hspace{0.01in}R}}).
\end{align}
There is also an additional gauge singlet superfield,  $M,$ that is a bi-fundamental of the flavor symmetry:
\begin{align}
M \sim ( \square_{N_{F\hspace{0.01in}L}}, \overline{\square}_{N_{F\hspace{0.01in}R}}).
\end{align}
The meson superfield $M$ may be thought of as a composite of electric quarks, with $M_{ij} = Q_i Q_j^c,$ whereas the magnetic quarks $q, q^c$ have no obvious expression in terms of the electric variables.

This infrared theory is IR free with $N_f > 3N.$ The metric on the moduli space is smooth about the origin, so that the K\"{a}hler potential is regular and can be expanded as
\begin{align}
K = \frac{1}{\beta} \Tr (|q|^2 + |q^c|^2) + \frac{1}{\alpha |\Lambda|^2} \tr |M|^2 + ...
\end{align}
The coefficients $\alpha, \beta$ are $\mathcal{O}(1)$ positive real numbers that are not precisely determined in terms of the parameters of the electric theory, and at best may be estimated by na\"{i}ve dimensional analysis \cite{Manohar:1983md, Georgi:1986kr, Luty:1997fk}; the size of these couplings will be relevant in our subsequent inflationary analysis.

After appropriate rescalings to obtain canonically-normalized fields $M, q, q^c$, the tree-level superpotential in the magnetic theory is given by
\begin{align}
W_m=y\,\tr\,q Mq^c-\mu^2 \tr\,M,
\label{eqn:msuperpot}
\end{align} 
where $\mu^2\sim m\Lambda$ and $y \sim \sqrt{\alpha}.$ Supersymmetry is spontaneously broken at tree level in this theory; the $F$-terms of $M$ are 
\begin{align}
F^\dagger_{M_{ij}} &= y\, q^{a\,i} q_{a}^{c \,j} - \mu^{2} \delta^{ij} 
\end{align}
which cannot vanish uniformly since $\delta^{ij}$ has rank $N_f$ but $  q^{a \,i} q_{a}^{c\,j}$ has rank $N_f - N_c < N_f.$ Hence supersymmetry is spontaneously broken in the magnetic theory by the rank condition. This SUSY-breaking vacuum lies along the quark direction,
\begin{align}
&\vev{M}_{\text{ssb}} =0   \hspace{1cm} \vev{q}_{\text{ssb}}=\vev{q^c}_{\text{ssb}}\sim \mu \unit_{N} 
\end{align}
with $\vev{F_M} \sim \sqrt{N_c} \, \mu^2$. A careful analysis of the one-loop contribution to the effective potential around this vacuum reveals that there are no tachyonic directions, and all classical pseudo-moduli are stabilized at one loop. The remaining fields not stabilized at one loop are goldstone bosons of the broken flavor symmetry, and remain exactly massless to all orders.

The theory also possesses supersymmetric vacua, in accordance with the non-vanishing Witten index of SQCD. For large values of the meson vev, the quarks become massive and may be integrated out below the scale $\vev{M}$; here the magnetic theory becomes pure $SU(N)$ super-Yang Mills.  This theory has a dynamically generated strong coupling scale, $\Lambda_m(M)$, given by
\begin{align}
\Lambda_m(M)  = M \left( \frac{M}{\Lambda}\right)^{\frac{a}{3}},
\end{align}
where $a=\frac{N_f}{N}-3,$ a strictly positive quantity when the magnetic theory is IR free.  Here, for simplicity, we have taken the meson vev to be proportional to the identity, i.e. $\vev{M}\sim M\unit_{N_f}$.  Gaugino condensation at this scale leads to an ADS superpotential \cite{Affleck:1983mk} for $M$.  Below the mass of the quarks this additional nonrenormalizable contribution obtains the form \cite{Affleck:1983mk} $W_{\text{det}}= N \left(y^{N_f} \frac{\det M}{\Lambda^{N_f-3N}}\right)^{\frac{1}{N}}$. The complete low-energy superpotential in the magnetic theory is then given by  
\begin{align}
W= N  \left( y^{N_f} \det M\right)^{\frac{1}{N}} \Lambda^{-a}  -\mu^2 \tr M + y \tr q Mq^c.
\label{eqn:superpot}
\end{align}
Interpreted physically, the parameter $a$ characterizes the irrelevance of the determinant superpotential. Larger values of $a$ correspond to $\frac{N_f}{N} \gg 1$ and an increasingly-irrelevant contribution from gaugino condensation.

The complete superpotential of the magnetic theory admits a SUSY-preserving solution with $F_M=0.$ However, this vacuum is very distant from the origin (and also the metastable vacuum) due to the irrelevance of the SUSY-restoring gaugino contribution to the superpotential; it lies in the meson direction at
\be
\vev{M}_{\text{SUSY}} = \frac{\mu}{y} \left(\frac{\Lambda}{\mu}\right)^\frac{a}{2+a} \unit_{N_f}  \hspace{1cm} \vev{q}_{\text{SUSY}}=\vev{q^c}_{\text{SUSY}}=0.
\ee
	The existence of supersymmetric vacua indicates the SUSY-breaking vacuum is metastable; in \cite{Intriligator:2006dd} it was shown that the metastable vacuum may be made parametrically long-lived. Explicitly, the bounce action corresponding to the nucleation of a bubble of true vacuum was found to be $S_4\sim \left(\frac{\Lambda}{\mu}\right)^{\frac{4a}{2+a}},$
which can be made arbitrarily large -- and the false vacuum long lived -- by taking $\mu\ll \Lambda$.
Ensuring that no transition to the supersymmetric minimum has occurred during the lifetime of the Universe (i.e., that the lifetime of the nonsupersymmetric universe exceeds 14 Gyr) places a constraint on the
theory \cite{Craig:2006kx}
\begin{align}
\label{Eq: Lifetime Constraint}
\frac{a}{a+2} \log \frac{\Lambda}{\mu} \gsim 0.73 + 0.003 \log\frac{\mu}{\tev} + 0.25 \log N.
\end{align}
This is a very weak constraint (amounting to $(\Lambda/\mu)^{a/(a+2)}\gsim 2$) , and one that is naturally satisfied by the hierarchy $\mu/\Lambda\ll 1$ required to keep the K\"{a}hler corrections under control. Although parametric longevity may be ensured by taking $\mu \ll \Lambda,$ notice also that the longevity bound also becomes weaker for large $a$ (correspondingly, $N_f \gg N$) since the operator creating the SUSY vacuum becomes increasingly irrelevant in this limit.

The parametric longevity of the metastable SUSY-breaking vacuum persists at finite temperature, and in fact finite-temperature effects have been shown to lead to preferential selection of the metastable-vacuum after high-scale reheating \cite{Craig:2006kx, Abel:2006cr, Fischler:2006xh}. Here we wish to consider a somewhat different scenario -- namely, that inflation itself may result from population of the metastable vacuum, with the mesonic scalar $M$ playing the role of the inflaton and the squarks $q, q^c$ serving as the waterfall fields of hybrid inflation. While it bears considerable resemblance to a multi-component version of supersymmetric hybrid inflation, this scenario is distinguished observationally by the unique features of the ISS model.     

\section{Mesonic inflation in SQCD: Rolling into the metastable vacuum}\label{sec:metastable}

Let us now consider the inflationary dynamics resulting from massive SQCD with $N_c < N_f < \frac{3}{2} N_c.$ At high energies the theory is characterized by the electric description, with superpotential (\ref{eqn:esuperpot}) and strong-coupling scale $\Lambda.$ Below the scale $\Lambda,$ the dynamics are described in terms of the IR-free magnetic variables with superpotential (\ref{eqn:msuperpot}). Given a random initial vev for the mesonic scalar $M,$ inflation may occur as the theory settles into the metastable SUSY-breaking vacuum.

\subsection{Slow-roll trajectory}

To analyze the inflationary dynamics, let us parametrize the inflaton trajectory by $M \sim \frac{\varphi}{\sqrt{N_f}} \unit_{N_f},$ which maximally respects the $SU(N_f)$ global flavor symmetry of the theory. Then the tree-level scalar potential near the origin (neglecting the determinant superpotential responsible for creating the SUSY vacuum) is given by 
\begin{align}
\label{eqn:treescalarpot}
V_0 = N_f \mu^4 - y \mu^2 \tr (q q^c) - y \mu^2 \tr (q^* q^{c*}) + \frac{y^2}{N_f} |\tr (q q^c)|^2\\
 + \frac{y^2}{N_f} |\varphi|^2 \left[ \tr(|q^c|^2) + \tr(|q|^2) \right]. \nonumber
\end{align}

After diagonalizing the resultant mass matrix, the masses of the squarks along this trajectory are given by $m_s^2 = \frac{y^2 |\varphi|^2}{N_f} \pm y \mu^2,$ while those of the quarks are $m_f^2 = \frac{y^2 |\varphi|^2}{N_f}.$ The one-loop effective potential for $\varphi$ is then 
\begin{align}
V_{eff}(\varphi) = N_f \mu^4 + \frac{N N_f y^2}{32 \pi^2} \left[ 2 \mu^4 \log \left( \frac{y^2 \varphi^2}{N_f \Lambda^2} \right) + \left( \frac{y \varphi^2}{N_f} - \mu^2 \right)^2 \log \left( 1 - \frac{N_f \mu^2}{y \varphi^2} \right) \right. \\
\left. + \left( \frac{y \varphi^2}{N_f} + \mu^2 \right)^2 \log \left( 1 + \frac{N_f \mu^2}{y \varphi^2} \right)
\right]. \nonumber
\end{align}
The squarks in this theory become tachyonic and roll off into the SUSY-breaking vacuum below $\varphi_c = \sqrt{\frac{N_f}{y}} \mu.$ Taking the parametrization $\varphi = x \varphi_c = \sqrt{\frac{N_f}{y}} \mu x,$ we have
\begin{align}
V_{eff}(x) = N_f \mu^4 \left \{ 1 + \frac{N y^2}{32 \pi^2} \left[ 2 \log \left( \frac{ y x^2 \mu^2}{\Lambda^2} \right) + \left(  x^2 - 1 \right)^2 \log \left( 1 - \frac{1}{x^2} \right) \right. \right. \\
\left. \left. \left( x^2 + 1 \right)^2 \log \left(1 + \frac{1}{x^2} \right) \right] \right \}. \nonumber
\end{align}
The slow-roll parameters for this trajectory are the same as those of supersymmetric hybrid inflation, namely
\begin{align}
\epsilon &\simeq \frac{y^5 N^2 M_P^2}{128 \pi^4 \mu^2} x^2 \left[ (x^2 - 1) \log \left( 1 - \frac{1}{x^2} \right) + (x^2 + 1) \log \left( 1 + \frac{1}{x^2} \right) \right]^2 \\
\eta &\simeq \frac{y^3 N M_P^2}{8 \pi^2 \mu^2} \left[ (3x^2 - 1) \log \left( 1 - \frac{1}{x^2} \right) + (3 x^2 + 1 ) \log \left( 1 + \frac{1}{x^2} \right) \right].
\end{align}

These slow-roll parameters naturally satisfy the slow-roll constraints $\epsilon, |\eta| \ll 1$ provided $x > 1.$ Up to flavor rotations, the inflationary story is as follows: Assuming arbitrary initial vevs for the scalar components of the meson superfield $M,$ with $\vev{M} > \frac{\mu}{\sqrt{y}},$ the scalar components of $q, q^c$ roll rapidly to $\vev{q} = \vev{q^c} = 0$ due to the effective mass terms arising from the meson vevs. Since SUSY is broken away from the supersymmetric vacuum, mass splittings between the squarks and quarks provides a gently-sloping potential for $M$ at one loop, driving its slow-roll evolution towards the origin. 

For arbitrary meson vevs, the scalar potential contains a constant contribution to the energy density of order $N_f \mu^4;$ however, this is only reduced to $N_c \mu^4$ in the metastable supersymmetry-breaking vacuum. In the context of supergravity, we assume some mechanism arises in the present era to cancel the cosmological constant in this vacuum.\footnote{As was pointed out in  \cite{Intriligator:2006dd}, adding a constant term to the magnetic superpotential so that the metastable vacuum has our observed vacuum energy makes the SUSY vacua anti-deSitter; this may lead to a suppressed tunneling rate due to quantum gravity effects, thereby preserving the metastability of the SUSY-breaking vacuum against first-order transitions to the AdS SUSY vacua.} While $M$ is slowly rolling, the effective energy density $N_f \mu^4$ drives inflation. Inflation continues until the diagonal components of $M$ reach the critical value, i.e.,
\begin{align}
\vev{M_c} = \frac{\mu}{\sqrt{y}} \unit_{N_f}.
\end{align}
Here the squarks $q, q^c$ become tachyonic, but only $N$ of the $N_f$ flavors of $q,q^c$ obtain nonzero vevs in rolling off to the supersymmetry-breaking vacuum. This is a consequence of the rank-condition breaking of SUSY in ISS models, wherein there are not enough independent degrees of freedom to cancel all the F-terms of $M.$ When $\vev{M} \leq M_c,$ $N$ flavors of the squarks $q,q^c$ become tachyonic and roll off into the SUSY-breaking vacuum, with 
\begin{align}
\vev{q} = \vev{q^c}^T = \left( \begin{array}{c}
\frac{\mu}{\sqrt{y}} \unit_N \\
0
\end{array} \right).
\end{align}
In this vacuum the remaining $N_c$ flavors are stabilized at one loop, and $\eta \sim 1;$ thus inflation comes to an end. Notice that $\eta \sim 1$ at this stage even for the components of $M$ that do not obtain large masses from the squark vev; the one-loop effective masses for the pseudo-moduli at this stage are sufficient to terminate slow-roll inflation. Here the $SU(N_f)$ flavor symmetry of the magnetic theory is broken by the squark vev to $SU(N) \times SU(N_c).$

\subsection{Slow-roll constraints}

It is natural to consider whether this inflationary trajectory can produce sufficient e-foldings of inflation. As with supersymmetric hybrid inflation, the value of $\varphi$ at $N_e$ e-foldings from the end of inflation is
\begin{align}
\varphi_e = \frac{\sqrt{N_e N}}{2 \pi} y M_P. 
\end{align}
Whereas conventional supersymmetric hybrid inflation enjoys significantly sub-Planckian initial vevs with suitably small values of the yukawa coupling, one might worry here that $y$ is not naturally small at energies close to $\Lambda.$\footnote{It is important to note here that the effective yukawa coupling $y$ appearing in the scalar potential (\ref{eqn:treescalarpot}) contains factors of wavefunction renormalization from the K\"{a}hler potential, and thus depends logarithmically on the energy scale.} We may estimate the size of $y$ at the strong-coupling scale using na\"{i}ve dimensional analysis; from  \cite{Cho:1998vc} we have $y \lsim \frac{4 \pi}{\sqrt{N N_f}}$ near $\Lambda.$ For sufficiently large $N_f,$ $y$ may be made naturally $\mathcal{O}(10^{-1})$ or smaller near the strong-coupling scale, and runs free in the IR. 

 The total displacement of the inflaton (the sum of displacements of the $N_f$ diagonal components of $M,$ defining the inflationary trajectory) is $\varphi.$ In order to obtain a total inflationary displacement $\frac{\sqrt{N_e N}}{2 \pi} y M_P,$  each component of $M$ need only be displaced by a distinctly sub-Planckian amount $\sim \sqrt{\frac{N_e N}{4 \pi^2 N_f}} y M_P.$ With the NDA estimate for the size of $y$ below the scale $\Lambda,$  this corresponds to a required displacement of each individual field by $\sim \frac{2 \sqrt{N_e}}{N_f} M_P,$ which may be made sufficiently small for large $N_f.$ Moreover, this guarantees $\vev{M} \ll \Lambda,$ rendering the magnetic description valid throughout the inflationary trajectory.

Unlike the case of supersymmetric hybrid inflation, recall that there exist additional vacua in the ISS theory -- supersymmetric minima created by gaugino condensation far from the origin. A supersymmetric minimum lies along the inflaton trajectory at
 \begin{align}
\vev{\varphi}_{\SUSY} \sim \sqrt{N_f} \,  \mu \, \left( \frac{\Lambda}{\mu} \right)^{\frac{a}{a+2}} y^{- \frac{a+3}{a+2}}.
\end{align}
For initial field values of $\CO(M_P),$ it is plausible for the inflaton to roll away from the origin and into the supersymmetric vacuum, rather than towards the origin along the slow logarithmic potential generated by one-loop effects. It is doubtful whether inflation will occur while rolling into the SUSY vacuum, since large corrections to the K\"{a}hler potential so far from the origin make slow roll implausible in that direction. 

Naturally, one would like to check the field value above which rolling into the SUSY vacuum is preferred. For $\varphi \gg \varphi_c,$ the logarithmic effective potential for an inflaton component rolling toward the origin goes as 
\begin{align}
V_{eff}(\varphi) \approx N_f \mu^4 \left[ 1 + \frac{N y^2}{32 \pi^2} \left( \log \left( \frac{y^2 \varphi^2}{\Lambda^2} \right) + \frac{3}{2}  \right) \right],
\end{align}
whereas the tree-level contribution from gaugino condensation that gives rise to the SUSY vacuum is
\begin{align}
V_{susy}(\varphi) = N_f \Lambda^4 \left| y \left( \frac{y \varphi}{\sqrt{N_f} \Lambda} \right)^{2+a} - \frac{\mu^2}{\Lambda^2} \right|^2.
\end{align}
The maximum of this potential $V_{eff} + V_{susy}$ lies around 
\begin{align}
\varphi_{max} \approx \sqrt{N_f} \, \mu  \, \left(\frac{\Lambda}{\mu} \right)^{\frac{a}{a+2}} y^{- \frac{a+1}{a+2}} \left( \frac{N}{\pi^2} \right)^{\frac{1}{2+a}}  2^{\frac{5}{2+a}} a^{1/a}.
\end{align}
The requirement that the initial value of the inflaton lie within the basin of attraction of the metastable SUSY-breaking vacuum, rather than the SUSY-preserving one, amounts to the condition $\varphi_e \ll \varphi_{max}.$ For large $a$ (i.e., $N_f/N \gg 1$,) this corresponds to the constraint
\begin{align}
\sqrt{\frac{N_e N}{4 \pi^2 N_f}} y^2 \frac{M_P}{\Lambda} \ll 1
\end{align}
Clearly, this constraint is most readily satisfied for a strong-coupling scale $\Lambda$ close to $M_P.$ However, it is instrumental to consider how large $\Lambda$ may be before imperiling the radiative stability of the Planck scale. As noted in \cite{ArkaniHamed:2005yv, Distler:2005hi}, in effective theories with a large number of species, radiative stability of Newton's constant gives rise to a constraint on the size of the cutoff. Coupling the magnetic theory to $\mathcal{N} = 1$ supergravity, the contribution to the effective Planck mass from light fields is of the form
\begin{align}
\delta M_P^2 \sim \frac{\Lambda^2}{16 \pi^2} \left( \frac{\partial^2 K}{\partial \phi_\alpha^\dagger \partial \phi_\beta} \right)^{-1} \left( \frac{\partial^2 K}{\partial \phi_\alpha^\dagger \partial \phi_\beta} \right) =  \frac{\Lambda^2}{16 \pi^2} \left( N_f^2 + 2 N N_f \right).
\end{align}
Radiative stability of Newton's constant therefore suggests $ \Lambda \lsim \frac{4 \pi}{N_f} M_P.$ Thus the natural value of $\Lambda$ -- as well as the viable value most likely to favor slow roll towards the metastable vacuum -- is $\Lambda \sim 4 \pi M_P / N_f.$ With this value of $\Lambda$ and the NDA estimate of $y,$ the condition (\ref{eqn:superpot}) becomes
\begin{align}
\sqrt{\frac{ 4 N_e}{N N_f}} \ll 1.
\end{align}

The late turnover in the potential owes largely to the considerable irrelevance of the determinant superpotential. This suggests that, for large $N_f$ and naturally Planck-scale $\Lambda,$ the turnover may be pushed sufficiently close to $M_P$ to guarantee that $\varphi$ rolls towards the origin rather than the SUSY vacuum. Notice this is compatible with $a \gg 1,$ a condition that likewise supports the parametric longevity of the metastable SUSY-breaking vacuum. 

\subsection{SUSY $\eta$ problem}

Thus far our discussion has been restricted to the case of global supersymmetry; naturally, one would like to extend the analysis to the case of local supersymmetry. As in the case of supersymmetric hybrid inflation, there is {\it no} $\eta$ problem at leading order provided an exactly canonical K\"{a}hler potential $K = \tr |M|^2 + \tr |q|^2 + \tr |q^c|^2.$ The supergravity scalar potential is of the general form
\begin{align}
V_s = e^{K/M_P^2} \left[  \left( \frac{\partial^2 K}{\partial \phi^\dagger_\alpha \partial \phi_\beta} \right)^{-1} \left( \pd{W}{\phi_\alpha} + \frac{W}{M_P^2} \pd{K}{\phi_\alpha} \right) \left(\pd{W^\dagger}{\phi_\beta^\dagger} + \frac{W^\dagger}{M_P^2} \pd{K}{\phi^\dagger_\beta} \right) - \frac{3}{M_P^2} |W|^2 \right].
\end{align}
Along the inflationary trajectory of the ISS model ($M = \frac{\varphi}{\sqrt{N_f}} \unit_{N_f}, \, q = q^c = 0$), coupling the theory to supergravity thus yields the scalar potential
\begin{align}
V_s(\varphi^\dagger, \varphi) = N_f \mu^4 \exp \left[ \frac{\varphi^\dagger \varphi}{M_P^2} \right] \left[ \left(1 + \frac{\varphi^\dagger \varphi}{M_P^2} + ... \right)^2 - 3 \frac{\varphi^\dagger \varphi}{M_P^2} \right] \\
= N_f \mu^4 \left( 1 + \frac{(\varphi^\dagger \varphi)^2}{2 M_P^4} + ... \right) \nonumber
\end{align}
provided an exactly canonical K\"{a}hler potential. All terms proportional to $|\varphi|^2$ cancel directly, preserving the small mass of the inflaton; the leading supergravity correction is $\mathcal{O}\left(|\varphi|^4\right).$

However, this convenient cancellation should be considered a fine-tuning. Any additional contributions to the K\"{a}hler potential beyond canonical terms -- which one would anticipate in the context of effective field theory  -- might be expected to generate the usual SUSY $\eta$ problem. A K\"{a}hler potential for the inflationary trajectory of the form 
\begin{align}
K(\varphi^\dagger, \varphi) = \varphi^\dagger \varphi + b \frac{ (\varphi^\dagger \varphi)^2}{\Lambda^2} + ...,
\end{align}
where $b$ is a dimensionless coefficient, yields a scalar potential for the inflaton with terms quadratic in $\varphi,$
\begin{align}
V(\varphi) = N_f \mu^4 \left( 1 + \frac{b \varphi^\dagger \varphi}{\Lambda^2} + ... \right).
\end{align}
These corrections arise both in the context of global and local supersymmetry, leading to an inflaton mass $m_{\varphi}^2 = b N_f \frac{\mu^4}{\Lambda^2} \gsim b H^2,$ which kills inflation for $b \sim \CO(1)$. 

It is natural to consider how finely-tuned the coefficients of higher-order terms in the K\"{a}hler potential must be in order to preserve slow-roll inflation. Consider the magnetic K\"{a}hler potential in terms of the rescaled variables. Generalizing the large-$N_f$ arguments of \cite{Cho:1998vc} to the case of $N \geq 1,$ higher-order corrections to the K\"{a}hler potential should obtain the form
\begin{align}
\delta K(M^\dagger M) =  \frac{b_1 (4 \pi)^2}{N^2 N_f^2 \Lambda^2} \tr \left( M^\dagger M \right) \tr \left( M^\dagger M \right)+ \frac{b_2 (4 \pi)^2}{N^2 N_f \Lambda^2} \tr \left( M^\dagger M M^\dagger M \right) + ...
\end{align}
where $b_1, b_2$ are $\CO(1)$ coefficients.
Upon rescaling to obtain canonically-normalized fields, this translates to a K\"{a}hler potential along the inflationary trajectory of
\begin{align}
K(\varphi^\dagger \varphi) =   |\varphi|^2 + \frac{16 \pi^2 b_1}{N^2 N_f^2 \Lambda^2} \left(|\varphi|^2 \right)^2 + \frac{16 \pi^2 b_2}{N^2 N_f^2 \Lambda^2} \left(|\varphi|^2 \right)^2 + ...
\end{align}
These corrections to the K\"{a}hler potential give rise to an inflaton mass term of the form $m_{\varphi}^2 = 16 \pi^2 (b_1 + b_2) \frac{\mu^4}{N^2 N_f \Lambda^2},$ and thus correction to the slow-roll parameter $\eta$ of order $\delta \eta \approx 16 \pi^2 (b_1 + b_2) \frac{M_P^2}{N^2 N_f^2 \Lambda^2} \sim \frac{b_1 + b_2}{ N^2}.$ In order for this contribution to the slow-roll parameters not to interfere with slow-roll inflation, the coefficients $b_1, b_2$ must satisfy
\begin{align}
b_1, b_2 \ll N^2.
\end{align} 
Here the condition on $b_1, b_2$ may be reasonably satisfied for $N \gsim 1.$ Although quadratic corrections to the inflationary potential arise as expected, the coefficients may naturally be small enough to preserve slow-roll inflation without additional fine-tuning.  It is worth emphasizing that this statement relies upon na\"{i}ve dimensional analysis estimates of corrections to the K\"{a}hler potential; such corrections are not explicitly known in terms of microscopic variables. Yet it is suggestive that the SUSY $\eta$ problem for inflationary theories involving supersymmetric gauge dynamics may be resolved by carefully considering the strong-coupling behavior of the K\"{a}hler potential near the cutoff.

\section{Inflationary predictions for ISS-flation}\label{sec:predictions}

In order to obtain inflationary predictions, it is useful to employ the formalism of \cite{Sasaki:1995aw} for a multi-component inflaton. There arise contributions to the spectral index and density perturbations from fluctuations both along the inflationary trajectory (parameterized by $\varphi$) and orthogonal to the inflationary trajectory. In order to account for orthogonal contributions, consider parameterizing $M$ by 
\begin{align}
M = \frac{\varphi}{\sqrt{N_f}} \unit_{N_f} + \phi
\end{align}
where $\phi$ is an adjoint of the flavor symmetry, corresponding to directions orthogonal to the inflationary trajectory. The amplitude of density perturbations in this case is given by 
\begin{align}
P_{\zeta}(\varphi, \phi)^{1/2} = \frac{1}{2 \sqrt{3} \pi M_P^3} \frac{V^{3/2}}{\sqrt{V_{\phi} V_{\phi} + V_{\varphi} V_{\varphi}}}
\end{align}
where, e.g., $V_{\varphi} = \pd{V}{\varphi}.$ Assuming initial fluctuations are all roughly of the same order, the contributions from orthogonal directions are equal to those from the inflationary trajectory, hence
\begin{align}
P_{\zeta}(\varphi, \phi)^{1/2} \simeq \sqrt{2} P_{\zeta}(\varphi)^{1/2}  \simeq  \sqrt{\frac{8 N_e N_f}{3 N}} \frac{1}{y} \left( \frac{\mu}{M_P} \right)^{2} \simeq \sqrt{\frac{N_e}{6 \pi^2}} N_f \left( \frac{\mu}{M_P} \right)^2.
\end{align}
Notice that the constraint guaranteeing the validity of slow-roll inflation in the ISS theory, $N_f/N \gg 1$ likewise increases the amplitude of density perturbations and allows a greater separation of scales between $\mu, M_P.$ Recall that the WMAP normalization is $P_{\zeta}^{1/2} = 4.86 \times 10^{-5};$ this corresponds to a constraint
\begin{align}
\frac{\mu}{M_P} \sim  7  \times 10^{-3} N_f^{-1/2}.
\end{align}
Neglecting the dependence on $N_f,$ we would again obtain the inference of GUT-scale $\mu \sim 10^{15}$ GeV. This is somewhat unfortunate for low-scale gauge mediation of supersymmetry breaking. Even in gravity-mediated SUSY-breaking scenarios, this would naively lead to gaugino and gravitino masses of order $m_{1/2} \sim m_{3/2} \sim \mu^2/M_P \sim 10^{12}$ GeV, far too high for weak-scale SUSY. The situation is helped somewhat by $N_f \gg 1,$ but only by a few orders of magnitude for even an absurdly large value of $N_f;$ obtaining $m_{1/2} \sim m_{3/2} \sim M_{weak}$ (and thus $\mu \sim 10^{10}$)  in a gravity-mediated scenario would require $N_f \sim 10^{10}.$ Indeed, Maintaining the hierarchy $\mu \ll \Lambda \simeq 4 \pi M_P / N_f$ in this scenario instead constrains $N_f \lsim 10^6.$ 

 The spectral index of the density perturbations in this case is given by
\begin{align}
n_s -1 =  - \frac{3 M_P^2}{V^2} \left( V_{\varphi} V_{\varphi} + V_{\phi} V_{\phi} \right) + \frac{2 M_P^2}{V} \left(V_{\varphi} V_{\varphi} + V_{\phi} V_{\phi} \right) \left( \frac{V_{\varphi \varphi}}{V_{\varphi} V_{\varphi}} + \frac{2 V_{\phi \varphi}}{V_{\phi} V_{\varphi}} + \frac{V_{\phi \phi}}{V_{\phi} V_{\phi}} \right) \\
\approx - 2 \frac{y^2 N M_P^2}{4 \pi^2} \frac{1}{\varphi_e^2} = - \frac{1}{N_e} 
\end{align}
Thus we find the spectral index of density perturbations to again be
\begin{align}
n_s \simeq 1 - \frac{1}{N_e} \approx 0.98.
\end{align}
 While this appropriately red spectrum is appealing, the high scale of $\mu$ required to match the observed amplitude of density perturbations -- even provided $N_f \gg 1$ -- makes conventional scenarios for SUSY-breaking using ISS dynamics unattractive. 
 
 Of course, a high scale of SUSY breaking is not necessarily fatal, since the approximate R-symmetry of the ISS theory may be a compelling setting for  the realization of split SUSY \cite{Arkani-Hamed:2004fb, Giudice:2004tc, Arkani-Hamed:2004yi}. Here $m_{3/2}$ is far too large for anomaly-mediated split SUSY, which requires $m_{3/2} \lsim  50 \TeV$ for TeV-scale gauginos. However, if we assume cancellation of the cosmological constant in the SUSY-breaking metastable vacuum in a post-inflationary era, it is reasonable to consider generation of gaugino masses directly from R- and SUSY-breaking as in \cite{Arkani-Hamed:2004fb}. Provided the F-term of the chiral compensator is engineered so that $F_\phi \ll m_{3/2}$ (as might be obtained in a suitable extra-dimensional construction), one readily obtains a gravitino mass $m_{3/2} \sim \sqrt{N_c} \mu^2 / M_P \sim 10^{13} \GeV$ with gaugino/Higgsino mass $m_{1/2} \sim m_{3/2}^3 / M_P^2 \sim \TeV.$ Alternately, one might imagine the persistence of an unbroken R symmetry in the low-energy theory, whose breaking by additional dynamics generates suitably small gaugino and Higgsino masses.
 
 A more attractive scenario for conventional gauge or gravity mediation would require further separation of the scales involved in SUSY-breaking and the generation of density perturbations. Such a scenario may arise naturally in the context of ISS models, where inhomogenous preheating results from the breaking of the global flavor symmetry at the end of slow-roll.
	
\section{Curvature perturbations from broken flavor symmetries}\label{sec:broken}

In \cite{Kolb:2004jm, Matsuda:2006ee, Matsuda:2007tr} it was observed that curvature perturbations may be generated by inhomogenous preheating due to the breaking of an underlying global symmetry at the end of slow-roll inflation. Quantum fluctuations generated during the inflationary era correspond to fluctuations in the initial conditions of the preheating phase. Whereas in the case of unbroken global symmetry these fluctuations in the initial conditions lead to background evolutions that are related by time translation, in the case of broken global symmetry they give rise to inhomogeneities in preheating efficiency and thereby generating curvature perturbations. The scale of these curvature perturbations depends on the dynamics of both the inflationary and preheating phases, and may readily constitute the main source of perturbations to the background metric.

The ISS model provides a natural context for the realization of this scenario, since the $SU(N_f)$ global flavor symmetry of the theory is broken at the end of slow-roll inflation in the SUSY-breaking vacuum. Contributions to the power spectrum of curvature perturbations arising from inhomogenous preheating may be significant and, moreover, admit a lower scale of $\mu$ suitable for weak-scale SUSY breaking.

\subsection{Inhomogenous preheating}
	
Let us first briefly review the general mechanism elucidated in \cite{Kolb:2004jm}. In a multi-component inflationary scenario where the inflaton $\vec \phi$ consists of many background fields $\phi_i$ related by a global symmetry, there may arise fluctuations of the $\phi_i$ both parallel and perpendicular to the inflaton trajectory. Fluctuations parallel to the direction of motion in field space correspond to adiabatic curvature perturbations of the sort generated in single-field inflationary scenarios, while fluctuations orthogonal to the direction of motion correspond to isocurvature perturbations. At the time $t_0$ between the end of the slow-roll inflationary era and decay of the inflaton, the value of the background inflaton will have acquired a spatial dependence due to quantum fluctuations, 
\begin{align}
\vec{\phi}(t_0, x) = \vec \phi_0 (t_0) + \delta \vec \phi(t_0, x).
\end{align}
The values of the background inflaton fields $\phi_i$ at the end of slow-roll inflation serve as initial conditions for the comoving number density $n_\chi$ of particles produced during preheating \cite{Kofman:1997yn}. If the global symmetry of $\vec \phi$ is unbroken, these fluctuations are all related by symmetry transformations, and do not lead to fluctuations in preheating efficiency. If the global symmetry is broken before the preheating phase, however, these fluctuations are no longer related by simple transformations, and inhomogeneities in preheating may ensue. In this case, fluctuations in $\vec{\phi}(t_0, x)$ result in fluctuations of $n_\chi.$ Since the energy density generated during preheating is proportional to the comoving number density, $\rho_\chi \propto n_\chi.$ Assuming non-adiabatic pressure perturbations during the preheating phase may be neglected, curvature perturbations during preheating can thus be expressed in terms of the number density perturbation,
\begin{align}
\label{eqn:zetaeqn}
\zeta \equiv \psi - H \frac{\delta \rho_\chi}{\dot \rho_\chi} \approx \alpha \frac{\delta n_\chi}{n_\chi}.
\end{align}
Here we have taken the spatially flat gauge; the proportionality constant $\alpha$ depends on the redshifting of the preheating particle. Now fluctuations in $\vec{\phi}$ parallel to the direction of motion in field space lead to initial conditions for $n_\chi$ related by time-translation; the resultant adiabatic curvature perturbations could be gauged away by a suitable choice of slicing and threading. Fluctuations in $\vec{\phi}$ perpendicular to the direction of motion may not be gauged away in an analogous manner, and lead to observable fluctuations in $n_\chi.$ In this manner, isocurvature perturbations are converted into adiabatic perturbations through inhomogeneities in preheating efficiency. From Eqn. (\ref{eqn:zetaeqn}) we may estimate the curvature perturbations arising from inhomogenous preheating efficiency,
\begin{align}
\zeta \approx \alpha \pd{\ln(n_\chi)}{\phi_\perp} \delta \phi_\perp
\end{align}
where $\delta \phi_\perp$ denotes fluctuations perpendicular to the inflationary trajectory {\it during preheating}. It is crucial to note that the amplitude of quantum fluctuations $ \delta \phi_\perp$ is determined during the slow-roll inflationary phase, while $\alpha \pd{\ln(n_\chi)}{\phi_\perp} $ comes from the details of the preheating process. The resultant power spectrum and spectral index of the curvature perturbations from inhomogenous preheating are
\begin{align}
\CP_{\zeta}(k) &= \left[ \alpha \pd{\ln(n_\chi)}{\phi_\perp} \right]^2 \CP_{\delta \phi_\perp}(k) \\
n_s - 1 &= \frac{d \CP_\zeta}{d \ln k} = \frac{d \ln \CP_{\delta \phi_\perp}}{d \ln k}.
\end{align}

The key feature is that the power spectrum of these curvature perturbations is the product of orthogonal perturbations $ \CP_{\delta \phi_\perp}(k),$ determined during the slow-roll inflationary process, and an amplifying factor $ \left[ \alpha \pd{\ln(n_\chi)}{\phi_\perp} \right]^2$ from the inhomogenous preheating process after slow-roll inflation has ended. 

The fluctuations in $n_\chi$ may be readily calculated in models of instant preheating \cite{Felder:1998vq}. Here the inflaton $\phi$ is coupled to the scalar preheating field $\chi$ via an interaction $\lang_{\phi \chi} = - \half g^2 |\phi|^2 \chi^2,$ and $\chi$ is coupled to fermions $\psi$ by an interaction $\lang_{\chi \psi} = h \bar \psi \psi \chi.$ The process $\phi \rightarrow \chi \rightarrow \psi$ leads to efficient conversion of the energy stored in the inflaton into fermions. 

If the inflaton trajectory doesn't pass exactly through the minimum of the potential, but instead at a minimum distance $|\phi_*|,$ the preheat particles will be characterized by an effective mass $m_\chi^2(\phi_*) = g^2 |\phi_*|^2.$ Then the comoving number density of preheat particles is given by  \cite{Felder:1998vq}
\begin{align}
\label{eqn:preheatnum}
n_{\chi} = \frac{\left( g |\dot \phi_*| \right)^{3/2}}{8 \pi^3}  \exp \left[ - \pi \frac{m_\chi^2(\phi_*)}{g |\dot \phi_*|} \right].
\end{align}
The power spectrum of curvature perturbations may then be calculated from the dependence of $|\phi_*|$ on $\phi_\perp.$ Let us now consider the explicit realization of inhomogenous preheating in ISS-flation.

\subsection{Inhomogenous preheating in ISS-flation}

The ISS model possesses a large $SU(N_f)$ flavor symmetry that is broken to $SU(N_f) \rightarrow SU(N_c) \times SU(N)$ in the metastable SUSY-breaking vacuum. As this global symmetry is broken at the end of slow-roll inflation, prior to any preheating or reheating effects, it provides a natural context for the realization of inhomogenous preheating and the generation of curvature perturbations.

Consider the fate of components of the inflaton trajectory after slow-roll inflation has ceased, $\varphi \lsim \varphi_c.$ $N$ components obtain masses $m_1^2 \sim y \mu^2$ from the squark vev, while the remaining $N_c$ components are pseudo-moduli with vanishing tree-level masses. These pseudo-moduli instead obtain positive masses at one loop, which in  \cite{Intriligator:2006dd} were determined to be $m_2^2 \sim \frac{\log 4 - 1}{8 \pi^2} N y^2 \mu^2.$ 

Once the squarks roll off into the SUSY-breaking vacuum, the components of the inflaton feel a quadratic potential near the origin, but one that is much steeper for the $N$ components with tree-level masses than for the $N_c$ pseudo-moduli. Consider now the inflationary trajectory after slow-roll, parameterized by massive components $\varphi_1$ and pseudo-moduli $\varphi_2$:
\begin{align}
M = \left( \begin{array}{cc}
\frac{\varphi_1}{\sqrt{ N}} \unit_{N} & \\
 & \frac{\varphi_2}{\sqrt{N_c}} \unit_{N_c} \end{array} \right)
 \end{align}
 The potential near the origin is essentially quadratic, given by 
 \begin{align}
 V(\varphi_1, \varphi_2) \approx m_1^2 \varphi_1^2 + m_2^2 \varphi_2^2
 \end{align}
 with $m_1^2, m_2^2$ as above. Under the influence of the quadratic potential, the components $\varphi_1, \varphi_2$ roll to the origin. Recall that the inflaton is coupled to the squarks $q,q^c$ via an interaction 
\begin{align}
\lang_{M q q^c} = - \half \frac{y^2}{N_f} |\varphi|^2 \left(|q|^2 + |q^c|^2 \right)
\end{align}
and that there also exists a coupling of squarks to fermions of the form
\begin{align}
\lang_{q q^c \psi} = \frac{y}{\sqrt{N_f}} q \psi_\varphi \psi_{q^c} + \frac{y}{\sqrt{N_f}} q^c \psi_\varphi \psi_q.
\end{align}
These couplings give rise to preheating as the inflaton components roll through the origin, with the light components of the squarks $q,q^c$ playing the role of the preheat fields. Oscillations of the inflaton about the origin begin when $|M| < \frac{\mu}{\sqrt{y}},$ i.e., the amplitudes of oscillation are $|\varphi_1| = \varphi_{1,c} = \sqrt{ \frac{N}{y}} \mu, |\varphi_2| = \varphi_{2,c} = \sqrt{\frac{N_c}{y}} \mu.$ The velocities of the fields as they pass through the minimum of the potential are given by $|\dot \varphi_1|_{0} \approx m_1 \varphi_{1,c}$ and $|\dot \varphi_2|_{0} \approx m_2 \varphi_{2,c},$ respectively. As the fields begin to oscillate, $\varphi_1$ rolls much faster than $\varphi_2,$ with 
\begin{align}
\frac{|\dot \varphi_1|}{|\dot \varphi_2|} \propto \frac{m_1}{m_2} = \left( \frac{\log 4 - 1}{8 \pi^2} N y \right) ^{-1/2} \gg 1.
\end{align}
As such, when $\varphi_1$ rolls through the minimum of its potential and initiates preheating, $\varphi_2$ is still close to its initial amplitude. The velocity entering into the number density of preheat particles, Eqn. (\ref{eqn:preheatnum}), is dominated by $| \dot \varphi_1|,$ while the displacement at the minimum is set by the amplitude of $|\varphi_2|.$ To good approximation, then, $|\dot \phi_*| \approx m_1 \varphi_{1,c},$ $|\phi_*| \approx \varphi_{2,c},$ $m_\chi^2 \approx \frac{y^2}{N_f} \left( \varphi_1^2 + \varphi_2^2 \right),$ and the comoving number density of preheat particles is then given by
\begin{align}
n_\chi \approx \frac{ \left( \sqrt{N/N_f} y \mu^2 \right)^{3/2}}{8 \pi^3} \exp \left( - \frac{\pi y \varphi_{2}^2 }{\sqrt{N N_f} \mu^2} \right).
\end{align}

 The fluctuations in $n_\chi$ are dominated by those of $\varphi_2,$ so we have to leading order
\begin{align}
\frac{\delta n_\chi}{n_\chi} \approx - \frac{2 \pi y \varphi_2}{\sqrt{N N_f} \mu^2} \delta \varphi_2 \approx - \frac{2 \pi}{\mu} \sqrt{\frac{y}{N}} \delta \varphi_2.
\end{align}
The ensuing power spectrum of curvature perturbations arising from inhomogenous preheating in the ISS model is given by (see Appendix A for details)
\begin{align}
\CP_{\zeta}^{1/2} \approx \sqrt{\frac{y N_f}{48 N}} \frac{\mu}{M_P}.
\end{align}
 Whereas the adiabatic curvature perturbations arising during slow-roll inflation go as $(\mu/M_P)^2,$ those from inhomogenous preheating go as $\mu/M_P,$ allowing a greater separation of the inflationary and SUSY-breaking scale $\mu$ from $M_P.$ Matching the observed spectrum due to curvature perturbations of this type would entail 
\begin{align}
\frac{\mu}{M_P} \sim 9 \times 10^{-5} \cdot N^{3/4} N_f^{-1/4}
\end{align}
This separation of scales is nearly two orders of magnitude greater than considered previously; neglecting factors of $N, N_f$ it suggests $\mu \sim 10^{14}$ GeV, and further separation of scales is obtained at large $N_f.$ However, obtaining $m_{1/2} \sim M_{weak}$ in this scenario would still require $N_f \sim 10^{10},$ far too large to maintain the hierarchy $\mu \ll \Lambda \simeq 4 \pi M_P/N_f.$ Assuming preheating is relatively efficient, the curvature perturbations arising from inhomogenous preheating should dominate over those arising from the inflaton alone provided $N_f^{3/2} \lsim \frac{1}{10} \frac{M_P}{\mu}.$ For $\mu/M_P \sim 10^{-5},$ this is the case for $N_f \lsim 500.$

The spectral index of these curvature perturbations is
\begin{align}
n_s \approx 0.98,
\end{align}
preserving the red spectral index prediction from supersymmetric hybrid inflation. 

\subsection{Curvature perturbations from a separate preheating sector}

Alternatively, one might imagine coupling the inflaton to a separate preheating sector not itself embedded in the ISS model. Assuming $\varphi$ preferentially decays to preheat particles $\varsigma$ via an interaction of the form $\lang_{\varphi \varsigma} = - \half \lambda^2 |\varphi|^2 \varsigma^2,$ the resultant density perturbations are of the form
\begin{align}
\CP_\zeta^{1/2} \approx \frac{\lambda}{4 \sqrt{3}} \sqrt{ \frac{N_f N_c}{N y}} \frac{\mu}{M_P}.
\end{align}
Matching the observed curvature perturbations in this scenario would entail
\begin{align}
\frac{\mu}{M_P} \sim 9 \times 10^{-4} \cdot \lambda^{-1} N_f^{-5/4}.
\end{align}
Assuming $\lambda$ is perturbative, one may obtain weak-scale SUSY-breaking in this scenario from $N_f \sim 10^3;$ the hierarchy $\mu \ll \Lambda$ is automatically satisfied in this scenario. In this case, curvature perturbations from inhomogenous preheating would certainly dominate over those from slow-roll inflation. The spectral index is again $n_s \sim 0.98,$ independent of the preheating mechanism. This is perhaps the most compelling setting for inflation and weak-scale SUSY breaking in the ISS model.

\subsection{Non-gaussianities from inhomogenous preheating}

There is no {\it a priori} reason to expect the curvature perturbations arising from inhomogenous preheating to be entirely gaussian \cite{Kolb:2005ux}. Indeed, such perturbations may exhibit a significant degree of non-gaussianity and may serve to discriminate ISS-flation from more conventional models of supersymmetric hybrid inflation. 

The degree of non-gaussianity may be characterized by the non-linearity parameter $f_{NL}$ parameterizing the non-gaussian contribution to the Bardeen potential $\Phi,$
\begin{align}
\Phi = \Phi_G + f_{NL} \Phi_G^2.
\end{align}
Here $\Phi_G$ denotes the gaussian part. Assuming $|f_{NL}| > 1,$ $\Phi$ and $\zeta$ may be accurately related via
\begin{align}
\Phi = -\frac{3}{5} \zeta = - \frac{3 \alpha}{5} \frac{\delta n_\chi}{n_\chi}.
\end{align}
We may estimate $f_{NL}$ in ISS-flation by expanding $\delta n_\chi / n_\chi$ to second order in $\delta \varphi_2;$ for instantaneous preheating within the ISS sector this yields
\begin{align}
\frac{\delta n_\chi}{n_{\chi}} \simeq \frac{-2 \pi y \varphi_2}{\sqrt{N N_f} \mu^2} \delta \varphi_2 - \half \left( \frac{2 \pi y}{\sqrt{N N_f} \mu^2} - \frac{4 \pi^2 y^2 \varphi_2^2}{N N_f \mu^4} \right) (\delta \varphi_2^2)
\end{align}
As such, the non-linearity parameter is given by
\begin{align}
f_{NL} \simeq \frac{10}{3} \left( \frac{1}{2 \pi} \sqrt{\frac{N}{N_f}} - 1 \right)
\end{align}
Given $N/N_f \ll 1,$ this suggests that the non-gaussianities of the inhomogenous preheating curvature perturbations in this model are well approximated by
\begin{align}
|f_{NL}| \approx \frac{10}{3}.
\end{align}
A similar estimate is obtained for the case of a separate preheating sector, although in both cases it should be emphasized that this is only a rough approximation. This result corresponds well with intuition developed for non-gaussianities in inhomogenous preheating scenarios with small global symmetry breaking  \cite{Kolb:2005ux}; $\CO(1)$ global symmetry breaking, as in the case of ISS-flation, leads to a relatively small degree of non-gaussianity. It is interesting to note, however, that this degree of non-gaussianity is above the observational limit $f_{NL} \sim 1$ and close to the anticipated $|f_{NL}| \sim 5$ sensitivity of Planck and other future tests of the angular bispectrum \cite{ Bartolo:2004if}. Moreover, it is significantly larger than the negligible non-gaussianity expected for conventional models of hybrid inflation \cite{Enqvist:2004bk}. Given that ISS-flation and conventional supersymmetric hybrid inflation produce the same prediction of the spectral index, non-gaussianities arising from spontaneous symmetry breaking at the end of slow-roll inflation may serve as a key experimental discriminator.

\section{Conclusions}\label{sec:conclusions}

We have seen that inflation may naturally occur while rolling into the supersymmetry-breaking metastable vacuum of massive supersymmetric QCD. Although the combination of supersymmetry-breaking and inflation is more a matter of novelty than profundity, this scenario is particularly attractive in that it  contains a concrete UV completion of the inflationary sector and the potential for distinctive observational signatures. Successful slow-roll inflation in the ISS model requires a large number of flavors and relatively small magnetic gauge symmetry, as well as a natural hierarchy $m \ll \mu \ll \Lambda \lsim M_P.$ Although quadratic corrections to the inflationary potential arise in the presence of a non-canonical K\"{a}hler potential, the ensuing contribution to slow-roll parameters may be small enough to forestall the conventional SUSY $\eta$ problem. Moreover, the spontaneous global symmetry breaking that accompanies the end of slow-roll inflation in the ISS model may give rise to dominant curvature perturbations through inhomogenous preheating. Such perturbations may possess observable non-gaussianities, further distinguishing ISS-flation from its more conventional cousins. 

It is difficult to simultaneously obtain weak-scale SUSY breaking and the observed inflationary spectrum strictly from the dynamics of the ISS model; standalone ISS-flation would seem to favor split SUSY or other high-scale mediation. However, weak-scale SUSY-breaking using conventional gauge- or gravity-mediation is feasible if the primary contribution to primordial curvature perturbations arises from coupling to a separate preheating sector.  

It would be interesting to consider concrete realizations of split supersymmetry using ISS SUSY-breaking dynamics, given the high scale of SUSY-breaking required to match inflationary observables in the absence of inhomogenous preheating. Likewise, a more careful analysis of non-gaussianities arising from inhomogenous preheating may be useful in light of recent evidence for significant non-gaussianities in the WMAP 3-year data \cite{Yadav:2007yy}.

\section*{Acknowledgements}

I am especially indebted to Shamit Kachru for helpful discussions and comments on the manuscript. I would also like to thank Savas Dimopoulos and Jay Wacker for useful conversations. I acknowledge the hospitality of the International Centre for Theoretical Physics and the Simons Workshop at Stony Brook, where parts of this work were completed. This research is supported by an NDSEG Fellowship, National Science Foundation grant PHY-9870115, and the Stanford Institute for Theoretical Physics. 

\section*{Appendix A. Power spectrum from inhomogenous preheating }
The power spectrum of the curvature perturbations arising from inhomogenous preheating in the ISS theory is given by
\begin{align}
\CP_\zeta(k)  = \left[ \alpha \pd{\ln(n_\chi)}{\phi_2} \right]^2 \CP_{\delta \varphi_2} (k) = \frac{\pi^2 y}{4 N \mu^2} \CP_{\delta \varphi_2} (k) .
\tag{A.1}
\end{align}
The power spectrum of $\delta \varphi_2$ is set during inflation, and may be calculated accordingly. On superhorizon scales the amplitude of quantum fluctuations is given by
\begin{align}
|\delta \varphi_2(k)| \simeq \frac{H_k}{\sqrt{2 k^3}} \left( \frac{k}{a H} \right)^{\eta}
\tag{A.2}
\end{align}
and the resultant power spectrum is 
\begin{align}
\CP_{\delta \varphi_2}(k) \equiv \frac{k^3}{2 \pi^2} |\delta \varphi_2(k)|^2 = \left( \frac{H_k}{2 \pi} \right)^2 \left( \frac{k}{a H} \right)^{2 \eta}
\tag{A.3}
\end{align}
Evaluated at horizon exit, $k = aH,$ we arrive at the observable power spectrum
\begin{align}
\CP_{\delta \varphi_2} = \left( \frac{H}{2 \pi} \right)^2 \approx \frac{N_f \mu^4}{12 \pi^2 M_P^2}
\tag{A.4}
\end{align} 
Thus the power spectrum of curvature perturbations arising from inhomogenous preheating in the ISS model is given by
\begin{align}
\CP_{\zeta}^{1/2} \approx \sqrt{\frac{y N_f}{48 N}} \frac{\mu}{M_P}
\tag{A.5}
\end{align}

As for the spectral index of these curvature perturbation, we have
\begin{align}
n_s - 1 \equiv \frac{d \log \CP_{\zeta}}{d \log k} = \frac{d \log \CP_{\delta \varphi_2}}{d \log k} \approx 2 \eta \approx - \frac{1}{N_e}
\tag{A.6}
\end{align}

\bibliography{ISSInflationRefs}
\bibliographystyle{JHEP}

\end{document}